\documentstyle[11pt,aaspp4]{article}

\newcommand{\ea}{et al.}
\newcommand{\kms}{\mbox{km\thinspace s$^{-1}$}} 

\begin{document}

\title{The Global Kinematics of the Globular Cluster M15}
\author{G.A. Drukier, S.D. Slavin, H.N. Cohn, P.M. Lugger, and R.C. Berrington }
\affil{Department of Astronomy, Indiana University, Bloomington, Indiana}
\authoremail{drukier@pegasus2.astro.indiana.edu}

\author{B.W. Murphy}
\affil{Department of Physics and Astronomy, Butler University, Indianapolis, Indiana}

\and 

\author{P.O. Seitzer}
\affil{Department of Astronomy, University of Michigan, Ann Arbor, Michigan}

\begin{abstract}
We present velocities for 230 stars in the outer parts of the globular
cluster M15 measured with the Hydra multi-object spectrograph on the
WIYN telescope. A new Bayesian technique is used for analyzing the
data. The mean velocity of the cluster is $-106.9\pm 0.3$ km
s$^{-1}$. Rotation with an amplitude of $1.5\pm 0.4$ km s$^{-1}$ and a
position angle of $125\pm 19\arcdeg\ $ is observed and the model
including rotation is marginally favored over one without rotation.
The velocity dispersion decreases from the center out to about
7\arcmin\ and then appears to increase slightly.  This behavior is
strikingly different from the continued decline of velocity dispersion
with increasing radius that is expected in an isolated cluster.  We
interpret this as an indication of heating of the outer part of M15 by
the Galactic tidal field.\footnote{Table \protect{\ref{T:members}} in
this paper contains the velocities of the 230 stars deemed to be
members of the cluster. A table of 361 stars with velocities which are
non-members will not be published, this table and Table
\protect{\ref{T:members}} are both available electronically from the
Astronomical Data Center (ADC). The ADC's Internet site hosts WWW and
FTP access to the ADC's archives at the URL http://adc.gsfc.nasa.gov/}
\end{abstract}

\begin{table}
\dummytable\label{Table:Obs}
\end{table}
\begin{table}
\dummytable\label{T:members}
\end{table}
\begin{table}
\dummytable\label{T:variables}
\end{table}
\begin{table}
\dummytable\label{Table:Disp}
\end{table}
\begin{table}
\dummytable\label{T:rotation}
\end{table}

\section{Introduction}

\subsection{Theoretical Background}

A full understanding of globular cluster dynamical
evolution---including the core-collapse process and its
aftermath---requires that global dynamical models be fit to global
data sets.  In close analogy with the strong coupling between a
stellar core and envelope, there is a strong interaction between the
core and halo of a globular cluster.  For both stars and clusters, the
energy-transport rate in the outer parts determines the time-averaged
rate of energy generation in the core (Hut \ea\markcite{hut92} 1992).
In clusters, star-star gravitational scattering is the
energy-transport mechanism and hard binaries are the central energy
source.

Over the past few years, we have fit isotropic Fokker-Planck models to
the surface density and velocity dispersion profiles of several
collapsed-core clusters including M15 (Grabhorn \ea\markcite{gra92}
1992, Dull \ea\markcite{dull97} 1997), NGC~6624 (Grabhorn
\ea\markcite{gra92} 1992), and NGC~6397 (Drukier\markcite{druk95}
1995).  We have recently extended our Fokker-Planck approach to the
accurate treatment of an anisotropic velocity distribution (Drukier
\ea\markcite{druk97} 1997).  This will allow us to produce fully
global models that include the radial-orbit bias that develops in
cluster halos.  Fitting dynamical models depends crucially on the
availability of kinematic data, since surface-brightness or star-count
profiles alone do not strongly constrain such important parameters as
the mass-function slope or the total cluster mass.

In addition to evolving as a result of internal energy transport,
clusters also respond to external tidal perturbations resulting from
interactions with the Galaxy.  Such perturbations include shocks
caused by the cluster's passage by the bulge or through the disk, and
a steady tidal acceleration from the smoother halo potential.  In both
cases, there is energy input to the cluster, producing a ``heating''
effect, although the latter is usually treated as a boundary condition
causing the clusters to lose mass.  While primarily the energy is
directly deposited in the outer part of the cluster's mass
distribution, as was first demonstrated by Spitzer and
Chevalier\markcite{sc73} (1973) the evolution of the entire cluster is
affected.  In particular, they found that tidal shocking tends to
accelerate core collapse.

The extent of tidal heating and its effect on the evolution of a
cluster has been of continuing theoretical interest.  A particular
motivation is that tidal heating appears to be a major contributing
factor to the destruction of globular clusters (Aguilar, Hut, \&
Ostriker\markcite{aho88} 1988; Okazaki \& Tosa\markcite{ot95} 1995;
Capriotti \& Hawley\markcite{ch96} 1996; Gnedin \&
Ostriker\markcite{go97} 1997). Kundi\`c \& Ostriker\markcite{ko95}
(1995) have given the most recent theoretical analysis of the expected
heating of globular clusters by tidal shocks. They have shown that the
second-order, tidal-shock relaxation term, neglected in many previous
studies, is usually more important than the first-order term, both for
the impulse and adiabatic approximations.  Weinberg\markcite{w94}
(1994) demonstrated that the usual adiabatic approximation does not
apply in a three-dimensional potential, and this has been included in
the analysis of Kundi\`c \& Ostriker\markcite{ko95} (1995). From their
analysis, it appears that tidal effects are locally important as far
into a cluster as the half-mass radius.

Of particular relevance for comparison with the observations of the
global velocity dispersion profile of M15 presented here, are models
for cluster evolution that include tidal effects. What are needed are
numerical predictions to compare with our velocity distribution
observations. Unfortunately, most papers dealing with this have failed
to provide an analysis that may be directly compared with the usual
velocity dispersion profile determinations. One exception is Allen \&
Richstone\markcite{ar88} (1988).  This study, which only looked at the
first-order effect, found that the velocity dispersion increases for
the escaping stars found in the outermost part of a cluster. They
determined the tidal radius by the minimum in the velocity dispersion
profile. This tidal radius and the difference between the radial and
tangential velocity dispersion profiles depended on the nature of the
assumed cluster orbit. Oh \& Lin\markcite{ol92} (1992) present a
hybrid approach, which uses a Fokker-Planck approach for treating
internal relaxation (using the second order Fokker-Planck terms only)
and direct orbit integration for including the effects of the tidal
field.  However, they do not provide results for the evolution of the
velocity dispersion profile.

\subsection{Observational Background}

As the prototypical collapsed-core globular cluster, M15 has received
considerable observational attention.  Recent Hubble Space Telescope
(HST) imaging studies have probed the stellar distribution in the
central region with increasingly higher angular resolution (Lauer
\ea\markcite{lauer91} 1991, De Marchi \&
Paresce\markcite{dmp94}\markcite{dmp95} 1994, 1995, Guhathakurta
\ea\markcite{gu96} 1996, Sosin \& King\markcite{ks97} 1997).  The
current consensus of this work is that the surface density profile of
M15 has a central power-law form, with no clear evidence for a
resolved core.  The tightest upper limit on the core radius is
1\farcs5 (Sosin \& King\markcite{ks97} 1997).  These results have been
combined with ground-based surface brightness measurements (Lugger
\ea\markcite{lu95} 1995) and star counts (King \ea\markcite{k68} 1968)
to generate a global surface density profile for Fokker-Planck model
fits (Dull \ea\markcite{dull97} 1997).  The structure of the outermost
region of M15 has recently been studied by Grillmair
\ea\markcite{grill95} (1995).  They carried out two-color stellar
photometry using digitized Schmidt plates and used the color-magnitude
diagram to statistically correct for foreground contamination, thus
allowing them to map out the large-radius density profile of the
cluster.  They found an apparent tidal radius of 23\arcmin\ for M15,
based on a King\markcite{k66} (1966) model fit.  They also found an
excess population of cluster stars beyond this radius, which they
interpreted as a tidal tail.

A number of studies have been carried out over the past decade to
determine the velocity dispersion profile of M15 (Peterson, Seitzer,
\& Cudworth\markcite{psc89} 1989; Gebhardt
\ea\markcite{geb94}\markcite{geb97} 1994, 1997; Dull
\ea\markcite{dull97} 1997).  These studies have all concentrated on
the central region of the cluster.  The greatest radial offset for any
star in the Peterson \ea\markcite{psc89} (1989) sample is 4\farcm6;
the other studies surveyed the central $1-2'$ radius about the cluster
center.  Peterson \ea\markcite{psc89} (1989) and Dull
\ea\markcite{dull97} (1997) used long slit spectroscopy, while
Gebhardt \ea\markcite{geb94}\markcite{geb97} (1994, 1997) used
Fabry-Perot imaging spectrophotometry to scan the H$\alpha$ line.

The key finding by Peterson \ea\markcite{psc89} (1989) is that the
velocity dispersion profile of M15 rises rapidly towards the cluster
center.  While subsequent studies by Gebhardt
\ea\markcite{geb94}\markcite{geb97} (1994, 1997) and Dull
\ea\markcite{dull97} (1997) confirm the rising nature of the profile
within the central arc minute, these recent studies plus that of
Dubath, Meylan \& Mayor \markcite{dm94} (1994) obtained a central
velocity dispersions between 11 and 14~\kms, much less than the
25~\kms found by Peterson \ea\markcite{psc89} (1989).  As discussed by
Dull \ea\markcite{dull97} (1997), the global surface density profile
of M15 and the velocity dispersion profile out to about 4\arcmin\ is
well fitted by a Fokker-Planck model that contains a substantial,
centrally concentrated population of non-luminous remnants ---
presumably neutron stars.  Multi-mass, King-type models do not provide
as good a joint fit to the surface-density and velocity-dispersion
profiles (Dull \ea\markcite{dull97} 1997; Sosin \& King\markcite{ks97}
1997).

In this study, we present the first velocity information for the
outermost region of M15\@.  The 230 cluster members identified in our
sample primarily lie in the range of $1-16\farcm6$ from the cluster
center.  Thus, our work complements that of Gebhardt
\ea\markcite{geb97} (1997), who have presented velocities for 1534
stars that primarily lie within 1\farcm5 of the cluster center.  Our
median velocity accuracy is 0.3~\kms; the velocity accuracies for the
Gebhardt \ea\markcite{geb97} (1997) sample vary over a range of
$0.5-10~\kms$.

The following section describes our observations and presents our
velocity measurements. Our analysis technique, which is new for this
application, uses Bayesian statistical methods. These are described in
\S\ref{A:bayes}.  Section \ref{S:kinematics} gives our analysis of the
velocities of the members.  In the final section, we interpret our
results in the context of models for clusters evolving under the
influence of tidal effects.

\section{The Data}
\subsection{Observations}
All of the new observations discussed here were obtained using the
Hydra multi-fiber spectrograph on the 3.5m WIYN
telescope.\footnote{The WIYN Observatory is a joint facility of the
University of Wisconsin, Indiana University, Yale University, and the
National Optical Astronomy Observatories.} This instrument has 100
fibers which can each be placed within the 1\arcdeg\ diameter
observing field to 0\farcs 2 precision. The minimum fiber separation
is 36\arcsec, so while the central 0\farcm 5 of the cluster cannot be
efficiently observed, this instrument is ideal for observing the outer
region of globular clusters. We were able to observe up to 80 stars at
a time.  For all observations we used the echelle grating with an
order centered at 515 nm, in the neighborhood of the Mg b
lines. Approximately 20nm of the order was imaged on the 2048 pixel
long CCD for a dispersion of about 0.01 nm/pixel.  The comparison
source was a Th-Ar lamp.

Use of Hydra requires accurate positions for the stars to be observed.
The positions of the stars came from two sources, a $3\times 3$ mosaic
of Curtis-Schmidt frames in $V$ and $I$ and a list from K. Cudworth of
the stars on his M15 proper motion program. This list filled in the
center of the cluster which was too crowded on the Schmidt frames to
allow accurate photometry.  The positions of the stars were reduced to
the astrometric system of the {\it HST Guide Star Catalogue} and
proved, at the telescope, to be very reliable.

\begin{figure}[t]
\plotone{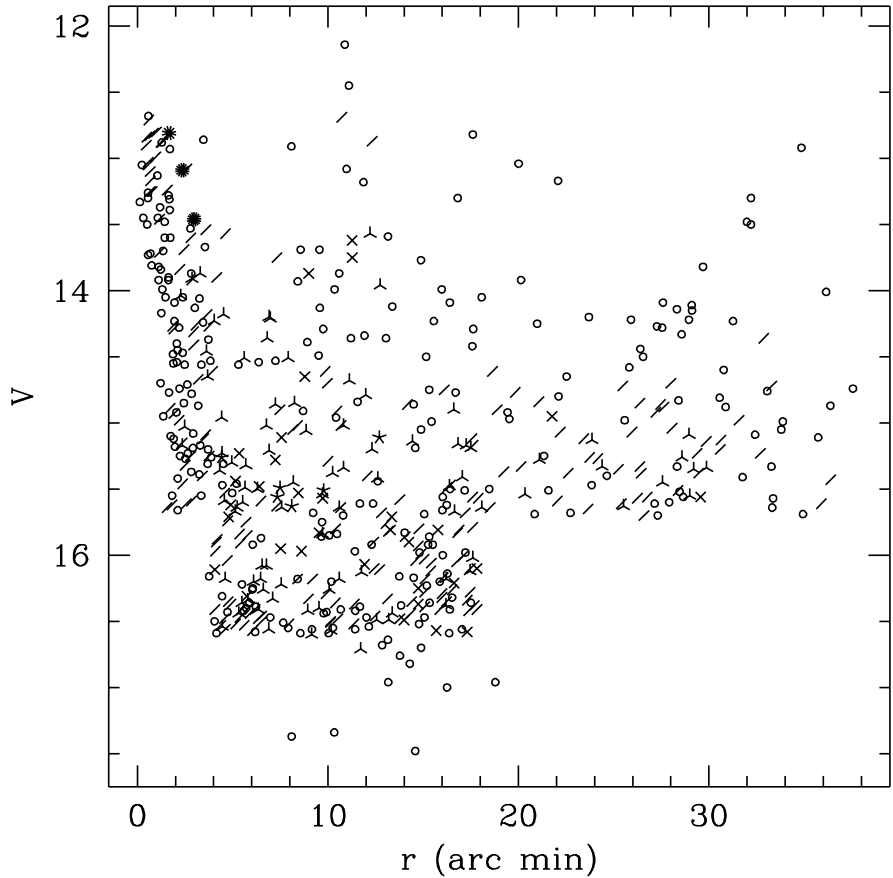}
\caption{Distribution of observed stars in radial position and $V$ magnitude.
The number of points on each symbol represents the number of observations. 
A small circle represents a single observation and a diagonal line two. 
\label{F:where}}
\end{figure}

The input list contained over 13,000 stars, the bulk of which are
non-members. Given our choice of spectral region, we restricted our
observations to stars on the giant branch. Our candidates were
therefore selected to lie in this region of the $V$ vs. $(V-I)$
color-magnitude diagram. We observed M15 during the course of three
observing runs, one each in May, June, and October 1996.
Table~\ref{Table:Obs} contains a log of the observations. The May 1996
run was in a sense experimental and took place before the full input
list was produced. We only had astrometry from the central Schmidt
frame as well as the positions of Cudworth's stars. Because of the
restricted region of candidates we could only observe of order 25
stars per Hydra set up. In June 1996 the full list was available and
we observed virtually all the stars on the giant branch brighter than
$V=15.7$ and outside about 4\arcmin.  In October 1996 we re-observed
all the known members between 4\arcmin\ and 18\arcmin, some of the
central stars, and an additional sample on the giant branch with
$15.7<V<16.6$ and $4\arcmin<r<17\arcmin$.  As the October run
progressed we did rough reductions of the spectra to weed out
non-members and to get additional spectra of apparent
members. Figure~\ref{F:where} shows the distribution of observed stars
in radius and magnitude. The diagram is a version of a `sunflower'
plot (Cleveland \& McGill \markcite{cm84}1984). The number of points
on each symbol represents the number of observations with the stars
represented by small circles being observed once and those by diagonal
lines twice. We obtained, in total, 1132 spectra of 591 stars in the
M15 field. Of these, 230 turned out to be members. Membership was
determined by velocity coherence and the strength of the absorption
lines as will be discussed further below.

\epsscale{0.75}

At least one of the stars K144 and K1040 were observed in each Hydra
configuration (a `setup').  The former proved to have a variable
radial velocity which changed by 1 km s$^{-1}$ between May and June
and 1.4 km s$^{-1}$ between June and October. The velocities for K1040
were much more consistent and its spectrum was used as the template
for the velocity determination by cross-correlation.

\subsection{Spectral Reduction}
The data were reduced using the {\it dohydra} reduction package in
IRAF\footnote{IRAF is distributed by the National Optical Astronomy
Observatories, which are operated by the Association of Universities
for Research in Astronomy, Inc., under cooperative agreement with the
National Science Foundation.}. Each observation was accompanied by one
or more 5-minute exposures of an incandescent lamp (a `flat') taken
with the fibers in the same configuration as the observations.
Generally, setups observed at the ends of the night had multiple flat
exposures, but, due to the overhead involved with flats and especially
with reconfiguring the fibers, usually single flat exposures were
done. There did not appear to be any disadvantage to using single
exposures since cosmic rays were not a great problem. The program
exposures and bracketing Th-Ar lamp exposures were extracted and then
divided by the extracted lamps. No sky subtraction was required owing
to the high-dispersion and the absence of the moon; all observations
were taken in dark time.  The wavelength calibration was done using 36
comparison lines.  The fifth-order dispersion solution generally had
RMS residuals of less than $10^{-4}$ nm or 0.05 km s$^{-1}$. During
dispersion correction the spectra were re-sampled into 2048
logarithmically spaced bins covering 20.7 nm in total.

Cosmic ray removal was done through the simple expedient of using the
IRAF {\it continuum} task to replace with the continuum fit all pixels
more than 4 standard deviations above the fit or 9 standard deviations
below the fit.  The latter was necessary due to the re-sampling of the
spectra during dispersion correction. The spline interpolation
required for re-sampling sometimes gave complimentary negative spikes
around large cosmic rays. Care was taken not to remove any genuine
absorption lines. Some artificial lines, arising in re-sampling but
falling under the nine-standard-deviation limit, may have been added,
but this would not have had a large effect on the resulting velocities
given the large number of real lines dominating the cross-correlation.

In 1996 May and October, multiple exposures were taken with each fiber
configuration. The resulting spectra were added together to produce
the final spectra for cross-correlation. The four exposures in the
October `Q' configuration were taken on two consecutive nights due to
clouds, but were nevertheless combined, since the faintest stars
lacked adequate flux in either pair. These were first shifted by a
velocity equivalent to the difference in the Earth's heliocentric
velocity relative to M15 between the two observations.

\subsection{Cross-correlation, the Velocity Uncertainties, and the Zero point }
\begin{figure}[t]
\plotone{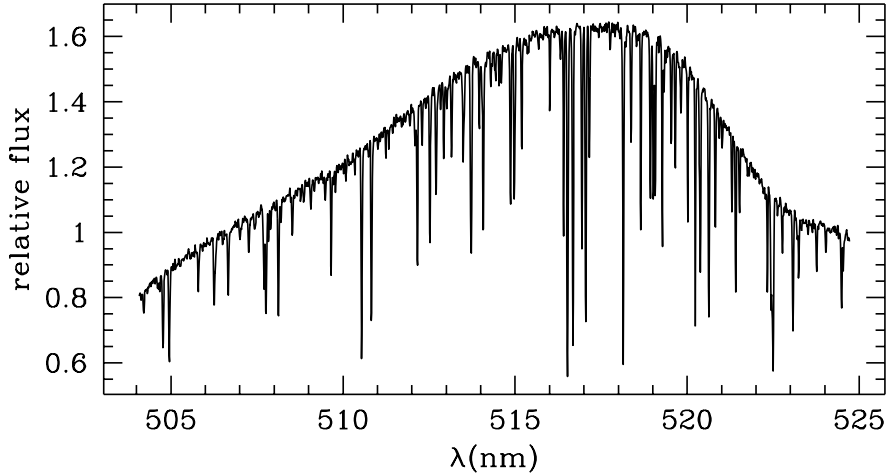}\caption{Template spectrum used for cross-correlations. This is a combination of
33 individual spectra of K1040 adjusted for heliocentric velocity differences. 
\label{F:template}}
\end{figure}

There were 33 individual exposures of K1040, totaling over 22 hours of
exposure time.  These were shifted to remove small differences in the
geocentric velocity and were summed to form the high signal-to-noise
template shown in Figure~\ref{F:template}. The stellar velocities were
measured relative to this template using the cross-correlation
technique of Tonry \& Davis \markcite{td79} (1979) encoded in the IRAF
{\it fxcor} task. This computes the cross-correlation function of the
Fourier transforms of the template and stellar spectra. The shift of
the cross-correlation peak from zero gives the velocity of the program
star relative to the template. This was computed for all the extracted
stellar spectra. We excluded from further analysis all spectra with
cross-correlation peaks less than 0.2. These spectra had low
signal-to-noise ratios and the velocities derived were often obviously
due to the selection of chance peaks in the cross-correlations, since
they often gave velocities of several 1000 km s$^{-1}$.

The uncertainties in the velocities, $\epsilon_v$, were determined
from the ratio, $R$, of the peak of the cross-correlation function to
the size of the random noise fluctuations. For each spectrum
$\epsilon_v = C/(1+R)$ (Tonry \& Davis \markcite{td79} 1979). The
constant $C$ depends on the number of resolution elements in the
observation and on the width of the cross-correlation function. In
practice the value of $C$ is established from the data.  We have used
the procedure of Pryor, Latham \& Hazen \markcite{plh88} (1988) to
calculate $C$. For all stars observed more than once in an observing
run we have computed the statistic
\begin{equation}
\epsilon = {\Delta v\over \left(\left(1+R_1\right)^{-2} +
\left(1+R_2\right)^{-2}\right)^{1\over 2}}.  
\end{equation} 
After weeding out possible variables we have 228 pairs of repeat
observations.  Using the Bayesian procedure described in
Section~\ref{A:bayes}, we have calculated the dispersion of the
distribution of $\epsilon$ assuming a zero mean. This gives $C=13.1
\pm 0.5$ km s$^{-1}$, which is the value used in computing the errors
in the velocity tables.

The velocity zero point was established using six, high
signal-to-noise exposures of the twilight sky, two taken in 1996 May
and four in 1996 October.  The fibers in these exposures had various
configurations.  The spectra in these exposures were extracted and
wavelength-calibrated in the usual way and these individual spectra
were cross-correlated against the K1040 template spectrum. The average
velocity for the spectra in each sky exposure was calculated. The
internal dispersions were between 0.10 and 0.24 km s$^{-1}$, and the
mean error in a single velocity was 0.6 km s$^{-1}$ using $C=13.1$ km
s$^{-1}$ as above.  The difference between the internal dispersion and
the individual uncertainties might suggest that the individual errors
are overestimated. On the other hand, the standard deviation of the
six mean velocities about their mean is 0.5 km s$^{-1}$ which is quite
consistent with the error estimate.  What this indicates is that while
the velocities of the spectra in an exposure can be measured
consistently, there are systematic differences between exposures as
well. Since there is agreement between the standard deviation of the
velocities in the six exposures and the errors of the individual
velocities, we can have some confidence that our value for $C$ and
hence for the uncertainties are correct. This is important for the
subsequent calculation of the cluster velocity dispersion. The zero
point is taken as the unweighted mean of the six mean velocities and
is $-99.4 \pm 0.2$ km s$^{-1}$. The velocity of K1040 itself is
derived in the same way as for the rest of the stars. Its mean
velocity with respect to the template is $ 0.04\pm 0.03 $ km s$^{-1}$.

\begin{figure}[t]\plotone{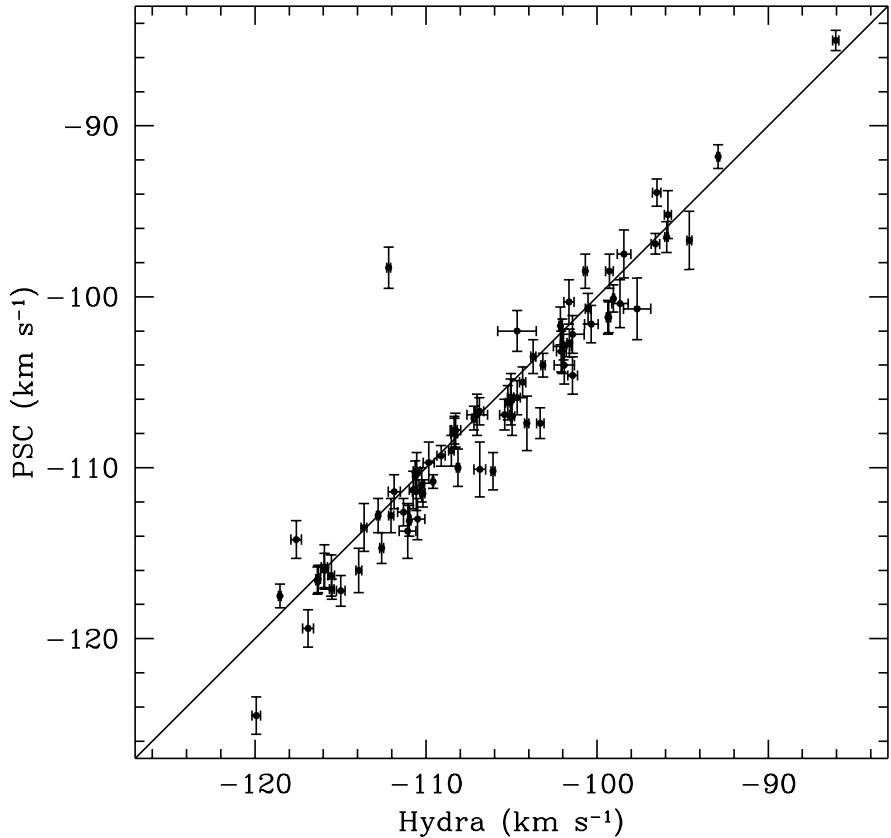}
\caption{Comparison of our velocities with those of Peterson et al. (1989).
The star with a large deviation is K673, a probable variable.
\label{F:psccomp}}
\end{figure}
\begin{figure}[t]\plotone{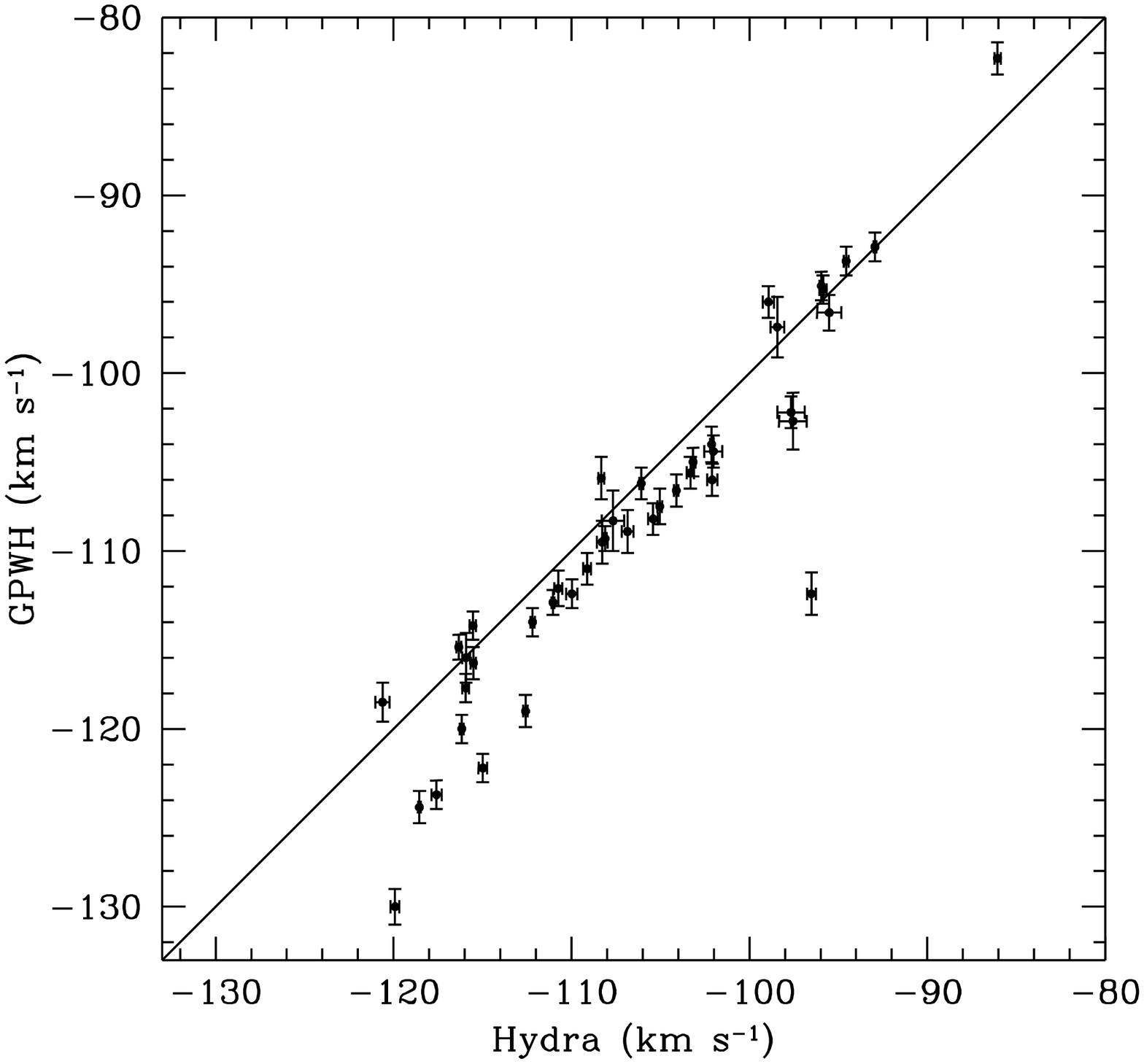}
\caption{Comparison of our velocities with the Fabry Perot velocities of Gebhardt
et al. (1994,1997).
\label{F:gebcomp}}
\end{figure}

As a final check we compare our velocities against the velocities of
stars in common with those observed by Peterson et
al. \markcite{psc89}(1989) and Gebhardt et al. \markcite{geb97}(1997).
Figures~\ref{F:psccomp} and \ref{F:gebcomp} compare the velocities of
the common stars.  There are 79 stars in common between our data and
Peterson et al.\markcite{psc89} (1989) and 42 between our data and
Gebhardt et al. \markcite{geb97}(1997), where we have only included
the new velocities reported by Gebhardt et al. The outlying star in
Fig.~\ref{F:psccomp} is K673 which is claimed by Gebhardt et
al. \markcite{geb94}(1994) to be a binary. The Peterson et
al. velocity is the most discrepant.  Most of the stars in
Fig.~\ref{F:gebcomp} lying off of the equality line are noted as being
variable in one or other of our studies. For four of the Gebhardt et
al. variables, we also have multiple observations and we confirm three
of them. One further possible variable is not confirmed. The apparent
zero-point shift is not statistically significant for the full sample,
but, for the non-variables, there is a correlation between declination
and velocity difference amounting to 5 \kms across the Fabry-Perot
field. Taking this into account, there appears to be a zero-point
difference of 0.9 \kms between the two data sets. These differences
may indicate a systematic problem with the Fabry-Perot calibration.

\begin{figure}[t]
\plotone{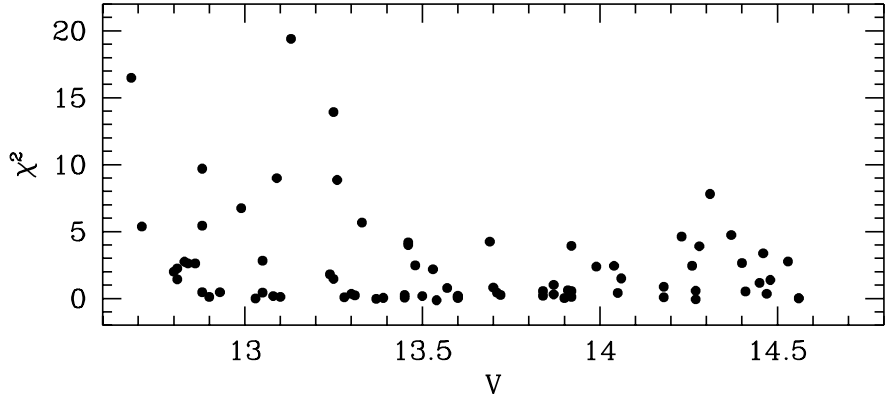}\caption{For the stars in common with
Peterson et al.  (1989) we show $\chi^2$ vs. magnitude.  Note the
extra variance apparent for stars brighter than $V=13.3$. This can be
attributed to the jitter seen in the velocities of bright giants.
\label{F:chicomp}}
\end{figure}

Calculation of the mean velocity differences for the full samples of
the two comparisons gives dispersions about the means which are
somewhat larger than expected if the adopted errors in each study are
correct.  Gebhardt et al. \markcite{geb97}(1997) note that their
measurement uncertainties are still not fully understood, but the
Peterson et al. uncertainties apparently are.  Thus, in
Figure~\ref{F:chicomp} we show, against $V$ magnitude, $\chi^2$ for
the two observations with respect to their mean for the comparison
with Peterson et al. sample. There is a difference in behavior for
stars on either side of $V=13.3$. There appears to be extra variance
for the stars brighter than $V=13.3$. This is probably due to the
velocity ``jitter'' seen for stars near the tip of the giant branch in
many clusters (Gunn \& Griffin \markcite{gg79} 1979; Mayor et
al. \markcite{may84}1984; Lupton et al. \markcite{lup87}1987; Pryor et
al. \markcite{plh88}1988).  Peterson et al. \markcite{psc89}(1989)
attribute the excess variance in their repeat observations to an
internal jitter of 0.88 km s$^{-1}$. The observations by Gebhardt et
al. \markcite{geb97}(1997) are generally not precise enough to see
this effect. If we restrict our comparison to stars with $V>13.3$ then
the difference distribution is consistent with unit variance. This
gives us confidence that our uncertainties have been calculated
correctly. The question of the ``jitter'' will be addressed further in
the next section.

\subsection{The Velocities}
\begin{figure}[t]\plotone{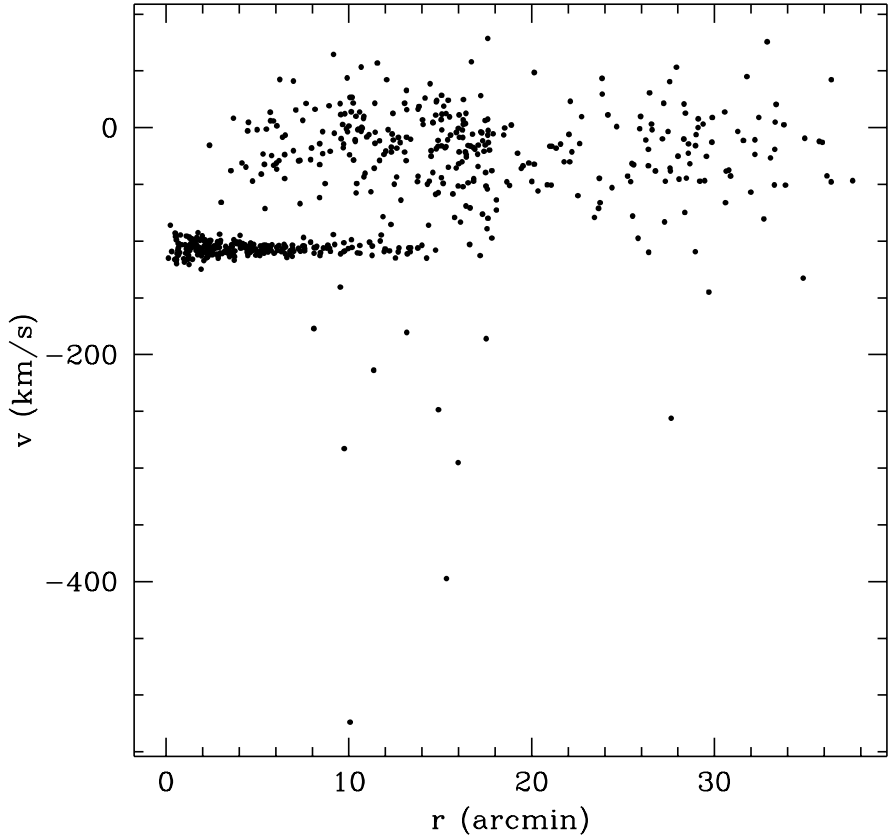}
\caption{Velocities for all the stars observed in this study plotted
against their radial positions. The cluster stands out as the clump
with velocity near $-107$ km s$^{-1}$. The stars around zero velocity
are disk stars, while the stars with very negative velocities probably
belong to the halo.
\label{F:allvsr}}
\end{figure}

In total we have measured 1132 velocities for 591 stars in the field
of M15.  Figure~\ref{F:allvsr} shows the measured velocities plotted
against distance from the cluster center. This was taken to be
$\alpha=21^{\rm h} 29^{\rm m} 58\fs 6$, $\delta=+12\arcdeg 10\arcmin
1\farcs 0$ (J2000) (Guhathakurta et al.\markcite{Gut96} 1996).
Generally, the cluster members stand out quite distinctly from the
non-members based on the roughly 100 km s$^{-1}$ velocity difference
between the cluster and the foreground disk stars. We selected the
cluster members by first excluding all stars with velocities greater
than 40 km s$^{-1}$ with respect to K1040. The spectra of all the rest
of the stars were examined individually to estimate the equivalent
widths of the Mg b lines.  These lines are a sensitive luminosity
indicator, but less so as a metallicity indicator (Geisler
\markcite{gei81} 1981), allowing us to distinguish between cluster
giants and field dwarfs.  All the stars with large equivalent widths
were rejected as being field Population I dwarf stars. A few of the
stars in the field region, those with the lowest velocities with
respect to K1040, were also examined to check for high-velocity
cluster members as have been seen in 47 Tuc (Meylan, Dubath, Mayor
\markcite{mdm91} 1991).  None was found.  We also examined the spectra
of 9 stars with velocities greater than 40 km s$^{-1}$ with respect to
K1040 that Cudworth (private communication) had assigned membership
probabilities in excess of 50\% based on their proper motions. All of
these had high Mg b equivalent widths.  There was also a population of
a dozen stars with velocities significantly more negative than that of
the cluster.  Several of these had very weak spectra, which precluded
estimates of the equivalent widths. Their velocities may be
suspect. Most of the others had equivalent widths larger than the
typical member, these are probable non-members. These have all been
rejected from the cluster sample.  These may be foreground halo
dwarfs.  There were also a few stars with velocities consistent with
being cluster members, but spectra too poor to estimate the equivalent
width. These have not been used in the subsequent analysis.

In Table~\ref{T:members} we have listed the stars deemed to be members
sorted by increasing right ascension.  The first two columns are our
coordinates for the stars with epoch 2000.  The third column contains
an identification to ease cross comparisons. For the stars from the
Cudworth's preliminary proper motion list, we have used his
identifications from K\"ustner\markcite{k21} (1921), Auri\`ere and
Cordoni\markcite{ac81} (1981), or from Sandage\markcite{san70}
(1970). These can also be used for cross-identification with Peterson
et al.\markcite{psc89} (1989). For the rest of the stars we have used
our own number from our master list of candidates. Further
identification numbers for cross referencing with Gebhardt et
al.\markcite{geb97} (1997) are in the penultimate column.  The fourth
column is a $V$ magnitude.  The remaining columns are the number of
observations for each star, the mean velocities, the uncertainties in
the velocities (exclusive of the uncertainty in the velocity-zero
point), and, for each star with more than one observation, the
probability that the $\chi^2$ of the differences of the observed
velocities about their mean is consistent with no variability. The
final column contains any notes.  The velocities of the non-members
are available from the authors.

We have compared our coordinates with the high precision astrometry of
the inner 2\arcmin\ of M15 by Le Campion, Colin, \&
Geffert\markcite{camp96} (1996). We find mean offsets (in the sense
ours--Le Campion et al.)  $\left<\Delta\alpha\right> = -0.68\arcsec
\pm 0.23\arcsec$, $\left<\Delta\delta\right> = +0.80\arcsec \pm
0.22\arcsec$. Since their typical error is around 0.07\arcsec, our
relative astrometry is good to about 0.2\arcsec.  The mean offsets
represent differences in the astrometric systems used in the
studies. We use the {\it HST Guide Star Catalogue} as our reference,
while Le Campion et al.\ use the FK5 system.  The positions in the
tables do not include the offset.  The zero-point for our photometry
has been determined by comparing stars in common between our list and
Cudworth's list, and shifting our magnitudes onto his system.

\begin{figure}[t]
\plotone{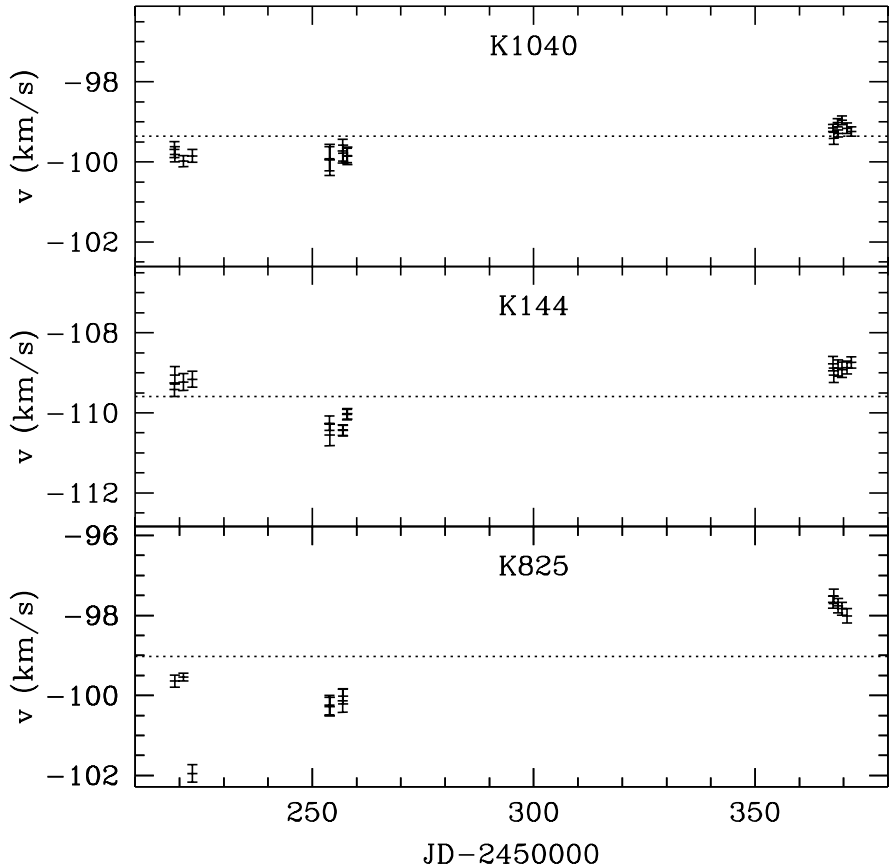}
\caption{Velocity curves for the standard stars K1040 and K144 and for
the suspected variable star K825. The errors shown do not include an
allowance for jitter.
\label{F:K825}}
\end{figure}

There are 17 stars in our member sample for which the probability of
no velocity variability is less than 1\%.  Of these stars, 12 are in
the upper one magnitude interval of the giant branch, here for
$V<13.6$, where it has long been suspected that the stars are subject
to an intrinsic ``jitter'' in their velocities of around 0.8 km
s$^{-1}$ (Gunn \& Griffin \markcite{gg79} 1979; Mayor et
al. \markcite{may84}1984; Lupton et al. \markcite{lup87}1987; Pryor et
al. \markcite{plh88}1988).  These include the two stars that we took
as our velocity standards, K144 and K1040.  Velocity curves for these
two stars, plus K825, are shown in Fig.~\ref{F:K825}. Both show
systematic variations between runs, but the velocities are consistent
if a further 0.8 km s$^{-1}$ is allowed for in the uncertainties.
There are another 12 stars with 2 observations, 6 of these are at two
epochs, which do not show any sign of this jitter.  So it is unclear
whether all bright giants suffer from this or what the time scale is
of this variation.  If we add 0.8 km s$^{-1}$ in quadrature to the
errors for all stars with $V<13.6$, then 3 of these stars remain
flagged as variables. These are marked by the note `b' in
Table~\ref{T:members}. Those with now consistent velocities are marked
`j'. The other five stars are fainter than this limit and have the
note 'v'. We provide the full set of velocities for the stars with `b'
and `v' flags in Table~\ref{T:variables}.  The star K673, which was
claimed to be a binary by Gebhardt et al.\markcite{geb94} (1994), only
appears to be variable if the jitter is not included. The two
observations in 1996 June and October differ by 0.8 km s$^{-1}$.  Of
the 8 stars we identify as velocity variables, 6 have only 2
observations, 4 of which are at two epochs. The star 5555 has a
velocity difference of 16 km s$^{-1}$ over a span of 3 days; star 6607
changes by 4.3 km s$^{-1}$ over the same interval. The most extensive
set of observations is for the star K825 which was noted to be
variable in our 1996 May run. We present the velocity curve in
Fig.~\ref{F:K825}. Most remarkable is a 2.4 km s$^{-1}$ velocity
change over the course of two days in our May observing run.

We will refer to the velocities without including the jitter factor as
our `A' sample.  This consists of 213 non-variables and 17
variables. If we include the 0.8 km s$^{-1}$ jitter factor in the
velocity uncertainties of all the bright stars, then this `B' sample
consists of 222 non-variables and 8 variables. Only the non-variables
in each sample will be used in estimating the mean, velocity
dispersion profile, and rotation. In the analysis below we will use
the A sample; the B sample gives very similar results.

\begin{figure}[t]
\plotone{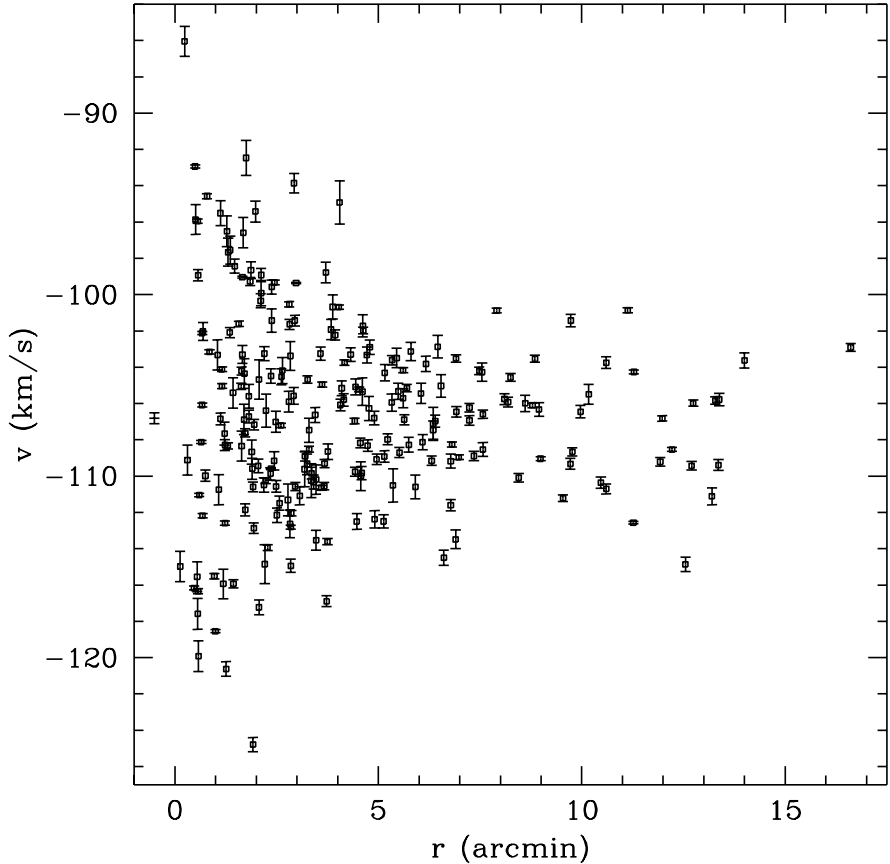}
\caption{Velocities of the members of M15 plotted against radius. The
velocity dispersion decreases as expected with radius until about
7\arcmin. The velocity dispersion then appears to increase again. The
point at negative radius indicates the inferred mean and its
uncertainty. In this case the probability distribution is symmetric.
\label{F:vvsr}}
\end{figure}

Figure~\ref{F:vvsr} shows our velocities plotted against radius. What
is most striking is the apparent increase in the velocity dispersion
beyond about 5\arcmin.  We will discuss this point in
\S\ref{S:kinematics} using the methods described in the next section.

\section{Bayesian Estimation}
\label{A:bayes}
\subsection{Introduction}
\label{A:intro}
We have a sample of velocities for members of the cluster, each with
its own uncertainty. What we wish to measure are the mean velocity of
the cluster and the velocity dispersion profile.  The mean must be
based on all of the velocities, but the velocity dispersion needs to
be measured as a function of radius, in radial bins, for example. In
addition, we often want to know if the observations indicate rotation
and whether the rotation signal is significant. There are several
approaches of increasing complexity which can be used to make these
various measurements.

Peterson et al.\markcite{psc89} (1989) used the simplest approach to
calculating the mean and simply took an unweighted average of their
120 stellar velocities and derived the uncertainty from the scatter
about the mean. This overstates the uncertainty since it will also
include the unknown velocity dispersion of the cluster. They
determined the velocity dispersion profile by dividing the sample into
bins and taking the dispersion about the mean of the velocities in
that bin as the local velocity dispersion. Again, the known and
variable uncertainties in the velocities are ignored. Furthermore, the
estimate of the cluster velocity based on the entire sample has also
been ignored in determining the local velocity dispersion. The
fractional error in the velocity dispersions was just taken as
$1/\sqrt{2N}$ for a bin of $N$ stars.

A more sophisticated approach is the maximum-likelihood method
described by Pryor \& Meylan \markcite{pm93}(1993). This assumes that
the velocity for each star is drawn from a normal distribution with
the standard deviation being the quadrature sum of the individual
velocity uncertainty and the cluster velocity dispersion. Standard
maximum-likelihood techniques result in equations for the mean and
dispersion which can then be solved numerically. Pryor \& Meylan also
give equations for the variances of the derived quantities.

Another algorithm which has been used to measure the velocity
dispersion profile utilizes the locally weighted scatter plot
smoothing (LOWESS) algorithm (Cleveland \& McGill
\markcite{cm84}1984). First the cluster mean is estimated by some
other method.  Then the velocity variance at each data point is
estimated in the following way, according to the application of this
algorithm by Gebhardt et al. \markcite{geb94}(1994).  The squared
deviations from the cluster mean are calculated. Then at each radius
for which one wants to measure the dispersion: (1) A straight line is
fit to these deviations as a function of radial position by weighted
least-squares.  The weights are the inverse squares of the differences
of the stars in radial position, measured from the cluster center,
with respect to the dispersion radius.  (2) The square root of the
fitted variance is taken as the velocity dispersion at that
radius. The uncertainties in the velocity dispersions are calculated
using a Monte Carlo method which assumes that the LOWESS dispersions
are correct, but using the observed velocity uncertainties for
individual stars. While this method does give a non-parametric
estimate of the velocity dispersion profile, it suffers from
calculating the mean and dispersions separately and from ignoring the
measurement errors in calculating the velocity dispersion. Thus all
velocities carry equal weight and the measurement of the velocity
dispersion can be biased by single, highly-uncertain points with large
deviations. To some extent this will be compensated for in estimating
the uncertainty in the velocity dispersions, and more recent
applications of the method have included the velocity uncertainty in
the weighting (Gebhardt, private communication), but the robustness of
this method has not been demonstrated.

Looking at the maximum-likelihood method from another perspective
brings us to the Bayesian methods we employ here. The results are
generally similar to those achieved with maximum likelihood, but there
are several advantages. The Bayesian methods give probability
distributions for the parameters, not just the most likely value and
its variance. The Bayesian methods naturally incorporate any prior
information on the values to be measured and also give the relative
likelihood of various models. Thus, for example, we can decide whether
a model including rotation is more or less likely than one without.

For discussions of the background of Bayesian analysis we refer the
reader to Bretthorst\markcite{b88} (1988) and Press
\markcite{p89}(1989). The classic source is Jeffreys
\markcite{j61}(1961). Jaynes\markcite{j83} (1983) presents this
material from a more modern perspective. Saha \& Williams
\markcite{sw94}(1994) have used these methods in a somewhat different
astronomical context.

In short, if we know the conditional probability of $A$ given $B$,
$P(A|B)$, then we can infer $P(B|A)$ using Bayes' theorem,
\begin{equation}
P(B|A) = {P(A|B) P(B) \over P(A)}.
\label{eq:bayes}
\end{equation}
In eq. (\ref{eq:bayes}) $P(B)$ is refered to as the {\em prior
probability} or the {\em prior}, it represents any information we have
on the values of the parameters before we look at our data.  $P(A|B)$
is the {\em likelihood}.  $P(B|A)$ is the {\em posterior probability};
it is this we are seeking to measure.  The final factor $P(A)$ is
called the {\em global likelihood} and is a normalization factor.

$A$ and $B$ can represent data, models, model parameters and so on. In
a Bayesian framework, the probabilities represent our state of
knowledge. This is unlike the more familiar ``frequentist'' viewpoint
which considers probabilities to just be the frequency with which
something occurs. Here it is perfectly meaningful to discuss the
probability of the parameters of a model having particular values. It
is just as meaningful to compare the probability of two models. What
is in question is not ``How often will such a model occur?'', but
``Given the previously known information and the data, what is the
probability that this model is correct?'' The prior represents our
state of knowledge before making the observation. We may know nothing
at all, in which case we would wish to choose as uninformative a prior
as possible. Alternatively, we may have previous measurements which we
are trying to refine. In this case, the proper prior is one which
represents the previous measurements. We will see examples of both of
these kinds of priors below. The appendices to Bretthorst
\markcite{b88}(1988) contain some useful comments on the choice of
priors.

If we have two different models, $M_1$ and $M_2$, and we wish to
choose between them, then the natural thing to look at is the
posterior odds ratio
\begin{equation}
{P(M_1|D)\over P(M_2|D)} = {P(D|M_1)P(M_1)\over P(D|M_2)P(M_2)},
\end{equation}
where $D$ is the data.  If the number or nature of the parameters in
the two models differ, it is important to keep all the normalization
terms in the priors and likelihoods.  If the odds ratio is greater
than unity then model $M_1$ is favored, otherwise model $M_2$ is more
likely. This will be used in discussing rotating and non-rotating
models for the M15 data.

Since the notion of a posterior-odds ratio is unfamiliar, we feel that
some further discussion is warranted to give a proper understanding of
our results.  Assume that $P_1(M_1|D)$ and $P_2(M_2|D)$ are the
probability distributions for the values of two parameters, $M_1$ and
$M_2$ given the data, and, further, assume that they are normal
distributions with means $\mu_i$ and dispersions $\sigma_i$, where
$i=1,2$. We can calculate the probabilities that $M_1 > M_2$ or that
$M_1<M_2$ and hence the odds of the two propositions. Define $k\equiv
(\mu_1-\mu_2)/\sqrt{\sigma_1^2+\sigma_2^2}$.  Then the odds of $M_1 >
M_2$ over $M_1<M_2$ are
\begin{displaymath}
{1 + \hbox{\rm erf}\left({k\over\sqrt{2}}\right) \over
1 - \hbox{\rm erf}\left({k\over\sqrt{2}}\right)}.
\end{displaymath}
If $\mu_1=\mu_2$, $k=0$ and the odds are even; the two propositions
are equally likely. For $k=1$, a $1 \sigma$ result, the odds are 5.3
to 1. That is, based on our prior knowledge and this data set, it is
5.3 times more likely that $M_1> M_2$ than that $M_1<M_2$.  If $\mu_1$
is $2 \sigma$ larger than $\mu_2$ the odds of $M_1 > M_2$ over
$M_1<M_2$ are 43 to 1. Similarly $k=3$ gives odds of 740 to 1.  If a
horse in a horse race has odds of 5 to 1 against its winning, you
wouldn't be surprised if it won. Similarly, if the posterior odds are
5 to 1 against $M_2$ being larger than $M_1$ you wouldn't be surprised
if it really was larger.  In general, we don't think of a $1\sigma$
result as being overly significant. If the odds were 740 to 1, we
would be surprised if the horse won, or if, with additional data,
$M_2$ proved to be larger than $M_1$.  But, we are similarly surprised
if a $3\sigma$ result proves to be wrong.  Nonetheless, it can be, and
long-shots do sometimes win.

\subsection{Velocity Models and Priors}
\label{S:Bmodels}
Here we will look at several models for a globular cluster data
sample, with different assumptions about the form of the velocity
dispersion profile and about whether the cluster is rotating.  The
first pair of models assume a single mean velocity for the sample and
then individual velocity dispersions for groups of stars binned by
radius. The sizes of the bins are variable, we assume only that the
velocity dispersion is the same for all stars in that radial bin. In
effect we assume that the velocity dispersion can be approximated by a
series of step functions. This is a general assumption of
binning. First we will consider a model without rotation.

The data sample consists of $N$ stellar velocities $v_i$ each with
uncertainty $\epsilon_i$. The model parameters are the mean velocity
$\bar{v}$ and the set of velocity dispersions ${\sigma_r}$ for the
$r=1,\ldots,M$ radial bins.  For a single observation we assume that
the likelihood of observing $v_i$ is given by
\begin{equation}
P(v_i|\bar{v},\sigma_r,\epsilon_i) = 
{1\over \sqrt{2\pi\left(\epsilon_i^2+\sigma_r^2\right)}}
\exp \left(-{\left(v_i-\bar{v}\right)^2\over 2 \left(\epsilon_i^2+\sigma_r^2\right)}\right).   
\label{eq:norot}
\end{equation}  
The likelihood of the whole data set $D$ is
\begin{equation}
P(D|\bar{v},\left\{\sigma_r\right\},\left\{\epsilon_i\right\}) =
\prod_i P(v_i|\bar{v},\sigma_r,\epsilon_i).
\end{equation}
Applying eq.(\ref{eq:bayes}) gives
\begin{equation}
P(\bar{v},\left\{\sigma_r\right\}|D,\left\{\epsilon_i\right\}) \propto 
P(D|\bar{v},\left\{\sigma_r\right\},\left\{\epsilon_i\right\}) 
P(\bar v) \prod_r P(\sigma_r) 
\end{equation}
for the posterior probability. We have ignored the normalization
factor $P(D)$ as we will only be concerned with the relative
probabilities of various parameter combinations and, for different
models of the same data, $P(D)$ is constant.

$P(\bar v)$ and $P(\sigma_r)$ are the priors for the mean velocity and
the velocity dispersion values. If we know nothing about the mean
velocity then the appropriate prior to use is a uniform prior. In one
sense this is a somewhat unusual probability distribution in that it
is not normalizable. In practice this is not usually a problem. If we
have a previous observation of the mean $\bar v_0 \pm \sigma_{\bar
v_0}$, then we could use a normal distribution
\begin{equation}
P(\bar v) = {1\over\sqrt{2\pi}\sigma_{\bar v_0}}\exp\left(-{(\bar v - \bar v_0)^2\over 2\sigma_{\bar v_0}^2}\right)
\end{equation}
for the prior. The appropriate uninformative prior for a scale
parameter such as the velocity dispersion $\sigma_r$ is the
Jeffreys\markcite{j61} (1961) prior $\sigma_r^{-1}$.  (For a
justification see Bretthorst\markcite{b88} 1988.)

An alternative model is one which allows for rotation in the data in
addition to the mean and dispersion as in the previous model. One
simple model is to assume a sinusoidal dependence of rotation velocity
on azimuthal angle with a single position angle, $\phi_0$ and
amplitude $A$. We replace eq.(\ref{eq:norot}) with
\begin{equation}
P(v_i,\phi_i|\bar{v},\sigma_r,\phi_0,A,\epsilon_i) = 
{1\over \sqrt{2\pi\left(\epsilon_i^2+\sigma_r^2\right)}}
\exp \left(-{\left(v_i-\left[\bar{v}+A\sin\left(\phi_i-\phi_0\right)\right]\right)^2
\over 2 \left(\epsilon_i^2+\sigma_r^2\right)}\right),   
\label{eq:rot}
\end{equation}  
where $\phi_i$ is the position angle for observation $i$.
The posterior probability is then 
\begin{equation}
P(\bar{v},\left\{\sigma_r\right\},\phi_0,A|D,\left\{\epsilon_i\right\}) \propto 
P(D|\bar{v},\left\{\sigma_r\right\},\phi_0,A,\left\{\epsilon_i\right\}) 
P(\bar v) P(\phi_0) P(A) \prod_r P(\sigma_r).
\end{equation}
The appropriate choice for the new priors is discussed in
Bretthorst\markcite{b88} (1988).  Based on his discussion we use
\begin{equation}
P(\phi_0) P(A) = {1\over 2\pi\delta^2} \exp \left(-{A^2 \over 2 \delta^2}\right).
\label{eq:prior rotation}
\end{equation}
The pseudo-dispersion $\delta$ is a {\em hyper-parameter}. This is a
case where we can not use an improper, i.e. non-normalizable, prior as
we want to compare the probabilities of the model with and without
rotation. The other parameters with improper priors are common to both
models, so their normalization divides out. The rotation parameters
are only in one model and so their normalization must be taken into
account for a proper comparison.  We can see in the raw data
(Figure~\ref{F:pa}) that any rotation for M15 must be small.  This
sets a limit on $\delta$. The value we choose for $\delta$ expresses
our estimate of the maximum amplitude of the rotation, the larger we
choose $\delta$ to be, the larger the rotation signal must be to be
considered significant with respect to the model without rotation.

As an alternative to a step-function for the velocity dispersion
profile, we can consider a power-law form:
\begin{equation}
\sigma(r) = \sigma_0 r^{\alpha}.
\end{equation} 
This is still parametric, but does away with binning. The likelihood for a single 
observation is now
\begin{equation}
P(v_i|\bar{v},\sigma_0,\alpha,\epsilon_i) = 
{1\over \sqrt{2\pi\left(\epsilon_i^2+\sigma_0^2 r^{2\alpha}\right)}}
\exp \left(-{\left(v_i-\bar{v}\right)^2\over 2 \left(\epsilon_i^2+\sigma_0^2 r^{2\alpha}\right)}\right),
\label{eq:norotpl}
\end{equation}  
and the posterior probability is 
\begin{equation}
P(\bar{v},\sigma_0,\alpha|D,\left\{\epsilon_i\right\}) \propto 
P(D|\bar{v},\sigma_0,\alpha,\left\{\epsilon_i\right\}) 
P(\bar v)  P(\sigma_0) P(\alpha).
\end{equation}
The new priors are given by $P(\sigma_0)=\sigma_0^{-1}$, the Jeffrey's prior we saw
above, and 
\begin{equation}
P(\alpha) = {1\over \sqrt{2\pi}\gamma} \exp \left(-{\left(\alpha-\alpha_0\right)^2\over
2\gamma^2}\right).
\label{eq:prior alpha}
\end{equation}
As in equation (\ref{eq:prior rotation}), $\alpha_0$ and $\gamma$ are
hyper-parameters.  Theory suggests $\alpha_0 = -0.5$, and we take
$\gamma = 1$.  The extension to the rotating case is straight forward.

\subsection{Metropolis Algorithm}
\label{S:mode}
We calculate the posterior probability distributions using the
Metropolis algorithm and follow the procedure in Saha \& Williams
\markcite{sw94}(1994). The basic algorithm is as follows. We'll refer
to a set of parameters as $\varpi$.  (1) Start with some values of the
parameters $\varpi$ and calculate the posterior probability
$P(\varpi|D)$.  (2) Pick, at random from a uniform distribution, a
possible change $\delta\varpi$ in the parameters and compute
$P(\varpi+\delta\varpi|D)$.  (3) If
$P(\varpi+\delta\varpi|D)>P(\varpi|D)$ then replace $\varpi$ by
$\varpi+\delta\varpi$ for the next iteration. If not, replace $\varpi$
by $\varpi+\delta\varpi$ with probability
$P(\varpi+\delta\varpi|D)/P(\varpi|D)$.  (4) Return to step (2) and
iterate.  The size of the trial changes must be large enough to ensure
that the entire range of acceptable values of the parameters are
covered, but be small enough that sufficient iterations accept
change. Given enough iterations, the distribution of accepted values
for the parameters converges to the posterior probability
distribution.

In principle we should save each set of parameters to look at the
full, multivariate distribution, but this would require too much
memory and require too many iterations if the number of parameters is
much larger than two. What we have done instead is to save the
distribution of each parameter individually. This is equivalent to
projecting the multivariate distribution along the axis of each
parameter, that is, to calculating all the marginal probability
distributions at the same time. Our estimator for each parameter is
then the mode of the individual probability distribution. The
posterior probability of the model is given by the overall probability
using these estimators of the modes.  With some multivariate
probability distributions, it is possible for the peaks of the
projected distributions to be significantly different from the global
peak of the distribution. To help guard against this we also keep
track during the iterations of the individual parameter sample giving
the highest posterior probability. In general this set of parameters
was close to, but not identical with the modes of the individual
distributions; the posterior probabilities were similar.  In practice,
the distributions from our data sets are unimodal and strongly peaked,
indicating that the modes of the projected distributions do represent
fairly the peak of the multivariate distribution.

These projected posterior probability distributions for the various
parameters were measured by counting the number of accepted parameter
values on a grid.  For a parameter $x$, $P_i\Delta x$ is the
probability of $x$ being between $x_i$ and $x_i+\Delta x$, where the
$x_i$ are the grid points.  We calculate the mode of $P(x|D)$ by
finding the maximum value of $P_i$, at $x_j$ say, and then using it
and the two bracketing values to find the parabola $y(x)$ for which
$\int_{x_i}^{x_i+\Delta x} y(x) dx = P_i\Delta x$ for each of
$i=j-1,j,j+1$.  The maximum of $y(x)$ is taken as the mode of the
distribution. Using this interpolation form, the mode lies at
\begin{equation}
x = x_{j-1} + {2P_{j-1} - 3 P_j + P_{j+1} \over
		P_{j-1}  - 2P_j + P_{j+1}} \Delta x 
\end{equation} 
The number of iterations and the grid spacings for each parameter were
chosen to ensure a smooth, well sampled distribution. Generally we
used of order $10^6$ samples to derive the posterior distributions.

\subsection{Examples}
\epsscale{0.5}
\begin{figure}[t]
\plotone{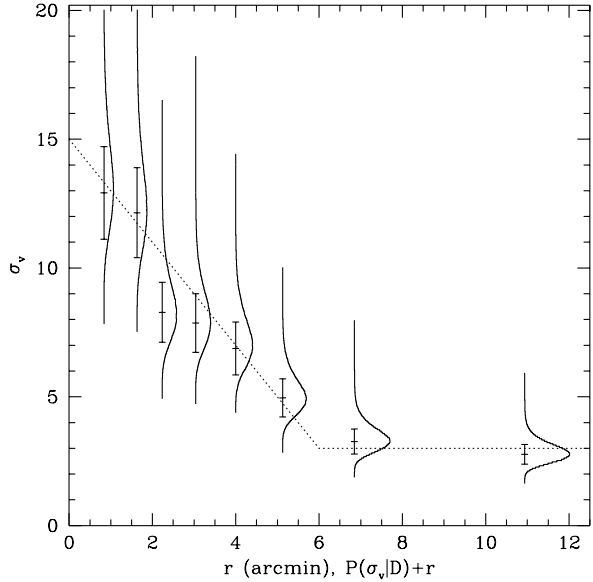}
\caption{Velocity dispersion profile for the first example. The
assumed velocity dispersion profile is shown as the dotted line. The
histograms (turn the figure sideways) show the probability
distributions for the velocity dispersion at that radius having the
given value. The probability distributions are offset to have their
zero levels at the mean radius of the star in the bin. This is the
same position as the vertical line of the error bars for the
accompanying points.  The points represent the modes of the
probability distribution and the error bars the symmetric region
containing 68.5\% of the probability. These can be thought of as
$1\sigma$ error bars. Note that the probability distribution is really
asymmetric; it is skewed to higher velocity dispersions.  There is
good agreement between the assumed profile and the estimated profile
from the artificial data.
\label{F:Bdisp}}
\end{figure}
\epsscale{0.75}

We conclude this discussion with a couple of examples to demonstrate
the method.  For first example, we draw a sample of stars with the
radii and errors of our `A' sample.  The velocities are drawn from a
distribution with a mean velocity of $-107$ km s$^{-1}$ and with a
velocity dispersion profile that decreases linearly to 6\arcmin, and
then is constant beyond that, similar to the observed velocity
dispersion profile We then use the same parameters as in
\S\ref{S:kinematics} to analyze this sample. The results are shown in
Figure~\ref{F:Bdisp}. The solid line is the assumed profile. The
estimates of the velocity dispersions are displayed in two ways on
this diagram: points with error bars and sideways histograms. The
first displays the values in a traditional manner. The data points
represent the mode of the distribution as discussed in
\S\ref{S:mode}. The error bars represent the symmetric region about
the mode over which the integrated probability is 0.685. Thus, they
are equivalent to 1$\sigma$ errors. Note, however, that a symmetric
probability integral is only one way of selecting the error
region. Any contiguous region containing 68.5\% of the probability
would be equivalent.  The agreement of the derived velocity dispersion
profile with that assumed is quite satisfactory and the inferred mean
velocity is $-107.0\pm 0.3$ km s$^{-1}$.

\begin{figure}[t]\plotone{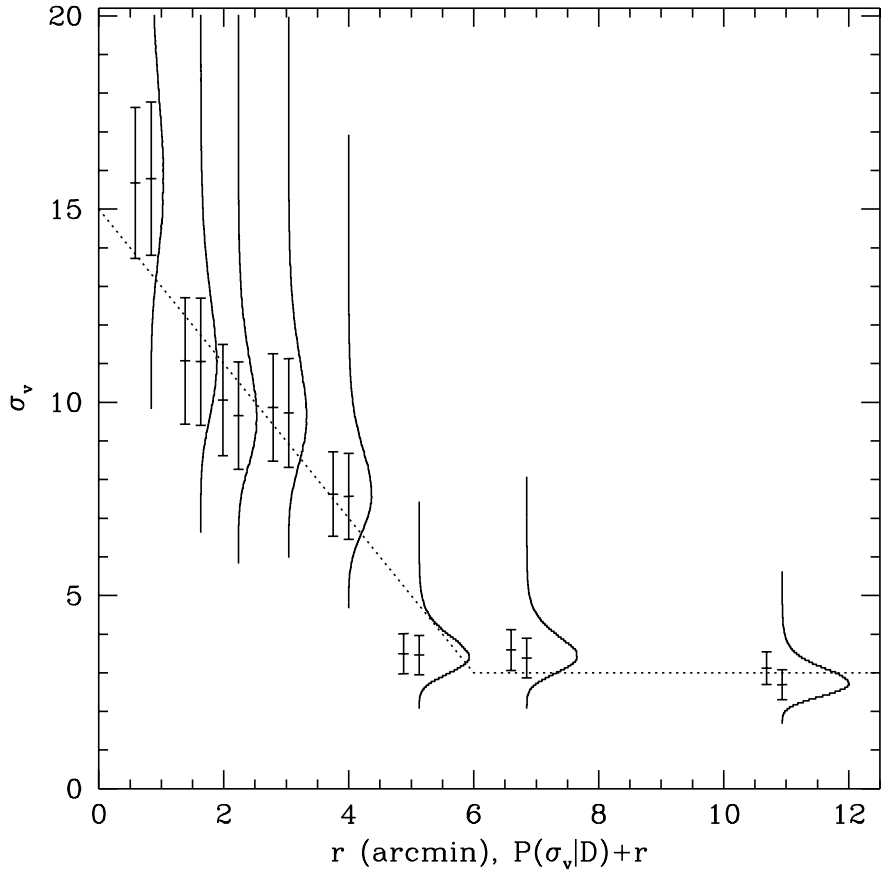}
\caption{Velocity dispersion profile for the second example. The
assumed velocity dispersion profile is shown as the dotted line. The
curves and points are as described in Figure~\protect{\ref{F:Bdisp}},
except that we also show the mode anderror bars for the solution
without rotation. These are offset to the left.
\label{F:Brot}}
\end{figure}

 For our second example we draw the stars as before, but also assume
the the entire cluster is rotating with amplitude 2 km s$^{-1}$ and a
position angle of 180\arcdeg. We analyzed this result with both the
rotating and non-rotating models.  The rotation is estimated to have
an amplitude of $1.7 \pm 0.5$ km s$^{-1}$ and position angle $176\pm
19\arcdeg$. The mean is $-106.7\pm 0.3$ km s$^{-1}$. These all agree
with the input model. The velocity dispersion profile is shown in
Figure~\ref{F:Brot}.  There is general agreement between the estimates
of the velocity dispersion and the assumed values. The model including
rotation is a more likely fit to the data with odds of 10 to 1 in
favor of rotation. This is only marginally significant, the data do
not strongly support an interpretation of rotation.  The sixth point
may look to be too low, but full consideration of the statistics shows
that this is not the case.  The advantage of having the full
probability distribution becomes apparent if we consider the odds
ratio for this point to be greater or less than the assumed value at
that radius. If we naively ignore the asymmetry and assume the
probability distribution to have the mean and standard deviation
shown, the difference is $2.5\sigma$ and the odds against the true
value being higher than the assumed value are 173 to 1. If we use the
measured probability distribution the odds are only 31 to 1, 5.5 times
less unlikely, and equivalent to a $1.3\sigma$ difference for normally
distributed errors.

\section{Kinematics}
\label{S:kinematics}
\subsection{Velocity dispersion profile}

We now apply to our observations the methods discussed in the previous
section.  We begin by assuming that the cluster is not rotating. We
divide the A sample into 7 bins of 26 stars each and an outermost bin
of 31 stars, and run it through the Bayesian analyzer. The resulting
mean velocity is $-106.9\pm 0.3 $ km s$^{-1}$ assuming a uniform prior
and $-107.3\pm 0.2 $ km s$^{-1}$ if we use the Gebhardt et
al.\markcite{geb97} (1997) mean and uncertainty ($-107.8\pm 0.3$ km
s$^{-1}$) as the prior. Our result in the latter case is just the
average of the two measurements, exactly as we would have
expected. The probability distribution of the mean velocity is well
represented by a normal distribution with the quoted mean and
dispersion.

\begin{figure}[t]\plotone{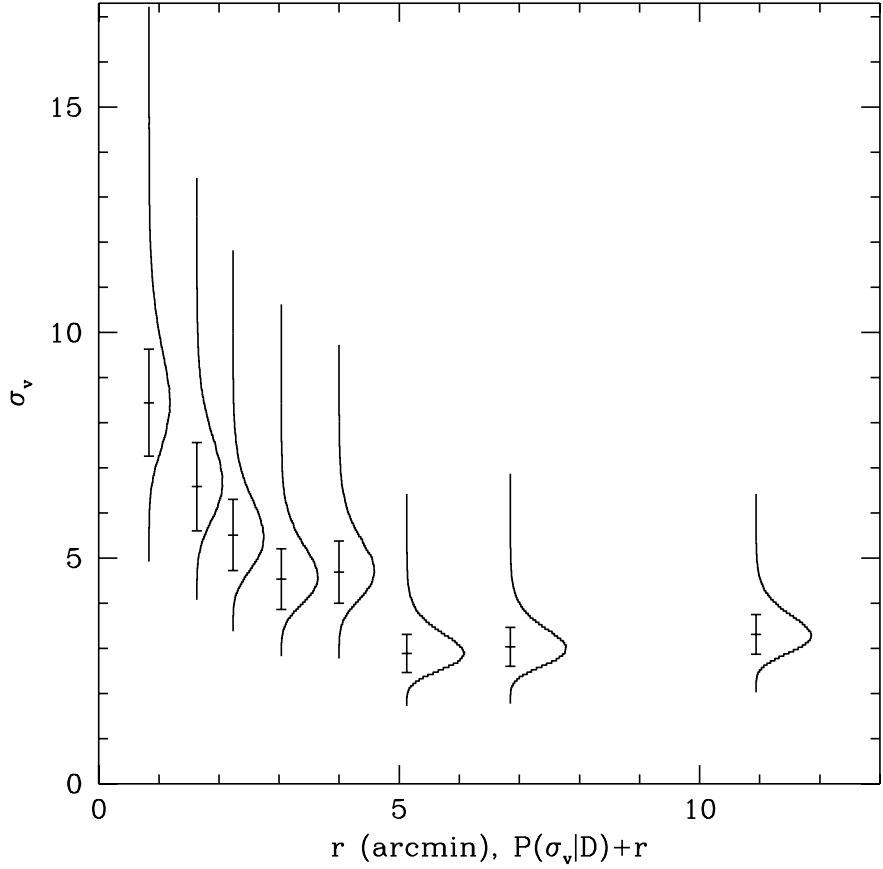}
\caption{The inferred velocity dispersion profile assuming no
rotation.  The curves and points are as described in
Figure~\protect{\ref{F:Bdisp}}.
\label{F:dispnorot}}
\end{figure}

Figure \ref{F:dispnorot} shows the resulting velocity dispersion
profile in the outer part of the cluster. If we use the B sample and
just add the `jitter' stars into the same radial bins, the results are
much the same. The points and curves are as described for the examples
in the previous section. The modes and the size of the symmetric error
region are listed in Table~\ref{Table:Disp}. The radius given in the
first column is just the mean radius of the stars in the bin.  The
histograms give the actual probability distributions for each velocity
dispersion. These can be seen more clearly by turning the figure
sideways. The zero level for the probability for each bin is the mean
radius of the stars in the bin as represented by the vertical stroke
of the error bars. It is clear that the probability distributions are
skewed in all cases to higher velocity dispersions. It is more likely
that the true dispersion is higher than the mode than less than
it. This is easy to understand. It is always possible that the
observed sample of stars lacks stars with large velocity differences
from the mean, even if the underlying velocity distribution allows for
such values.  On the other hand, the stars with the largest velocity
deviations in the sample put a much stronger lower limit on the
velocity dispersion.

With the exclusion of the point at 4\arcmin, the velocity dispersion
decreases with radius up to 5\arcmin. Beyond this radius, the modes of
the distributiqons increase again, confirming what is seen in the raw
velocities. Since we have computed the probability distributions, it
is straightforward to calculate the probability that one point is
larger than another, or, alternatively, the odds.  The probability
that the last point is higher than the next-to-last is 67\%, i.e. the
odds are roughly 2 to 1 in favor of the last point being higher. For
the last point and point at 5\arcmin, the probability that the last is
larger is 74\% giving odds of 2.8 to 1. (For an explanation of these
odds ratios, see the end of \S\ref{A:intro}.) While intriguing, these
results do not in themselves argue strongly for an increase in the
velocity dispersion.

\subsection{Rotation}
It could be argued that the increase in the velocity dispersion is due
to differences in the velocity tensor with radius.  If the velocities
are strongly tangentially anisotropic, the projected velocity
dispersion would be higher than if the velocities are isotropically
distributed. We discuss the case against tangential anisotropy in the
next section, but since rotation has been claimed in the past for M15,
here we look further at the special case of rotation. Gebhardt et
al.\markcite{geb97} (1997) find rotation with an amplitude of $2.1\pm
0.4$ km s$^{-1}$ and position angle $107\pm 10\arcdeg$ for their
overall sample.\footnote{Gebhardt et al.  define their position angle
as the velocity maximum. We define ours as the rotation axis with the
direction defined by right-handed rotation about it. We have
subtracted the required 90\arcdeg\ from their values to bring the
definitions into agreement.}  They also look at the variation with
radius and find changes in both the amplitude and position angle.

Before continuing on to look for rotation in our sample, we need to
address the question of ``perspective rotation'' arising from the
proper motion of M15.  This is discussed by Feast, Thackeray \&
Wesselink \markcite{ftw61}(1961) and more specifically with respect to
$\omega$ Centauri by Merritt, Meylan \& Mayor
\markcite{mmm97}(1997). Simply put, the projection of the cluster's
space velocity along the lines of sight to various parts of the
cluster results in an apparent rotation of the cluster, increasing
with distance from the cluster center and varying inversely with the
cluster distance.  For $\omega$ Cen, which is at 5.2 kpc and has a
total proper motion of 0.78 arcsec century$^{-1}$, the perspective
rotation is about 1 km s$^{-1}$ at 20\arcmin. M15 is at 10.5 kpc; the
value of its proper motion is disputed. Cudworth \& Hanson
\markcite{ch93}(1993) measured an absolute proper motion of
$\mu_\alpha \cos\delta = -0.03\pm 0.10 \arcsec/100a, \mu_\delta =
-0.42\pm 0.10\arcsec/100a$. Geffert et al. \markcite{gef93}(1993)
measured it to be $\mu_\alpha \cos\delta = -0.10\pm 0.14 \arcsec/100a,
\mu_\delta = -1.02\pm 0.14\arcsec/100a$.  More recently Scholtz et
al.\markcite{sch96} (1996) derived a value of $\mu_\alpha \cos\delta =
-0.01\pm 0.04 \arcsec/100a, \mu_\delta = +0.02\pm 0.03\arcsec/100a$.
None of these values agree, and the latest is effectively zero. Hence,
we will ignore the effects of perspective rotation here.

\begin{figure}[t]\plotone{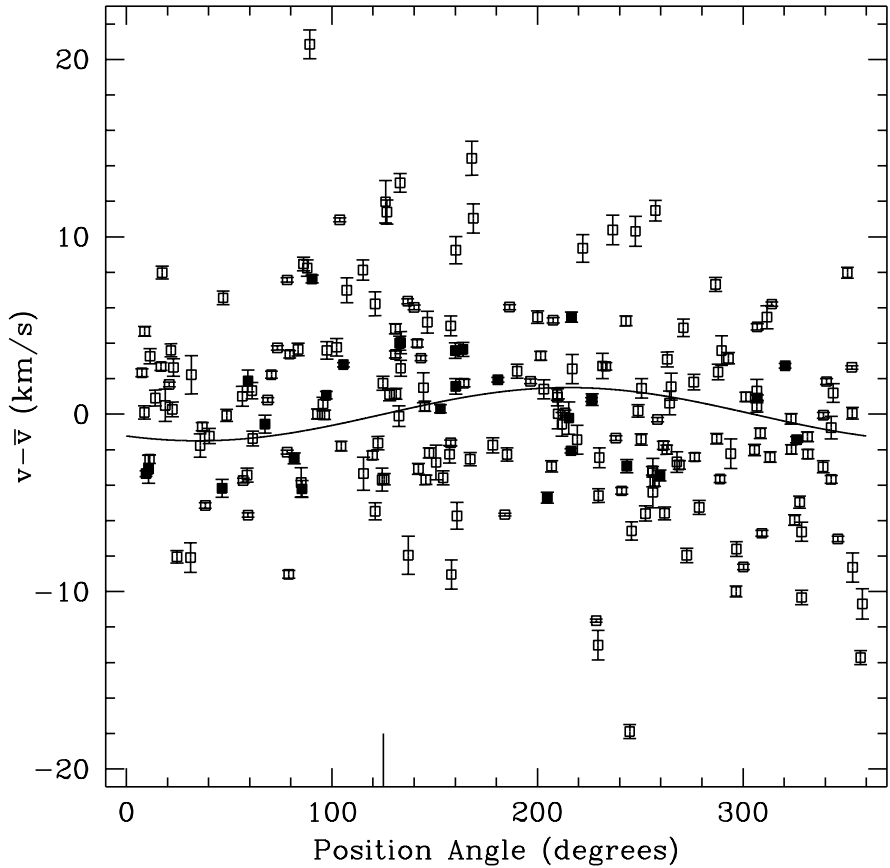}
\caption{Observed velocities, less the cluster mean, as a function of
position angle measured eastward from north. The inferred rotation is
shown as the curve. The position angle of the rotation axis is
indicated by the large tick mark. The stars in the 4\arcmin\ bin,
which has the most significant rotation, are shown as filled symbols.
\label{F:pa}}
\end{figure}

\begin{figure}[t]\plotone{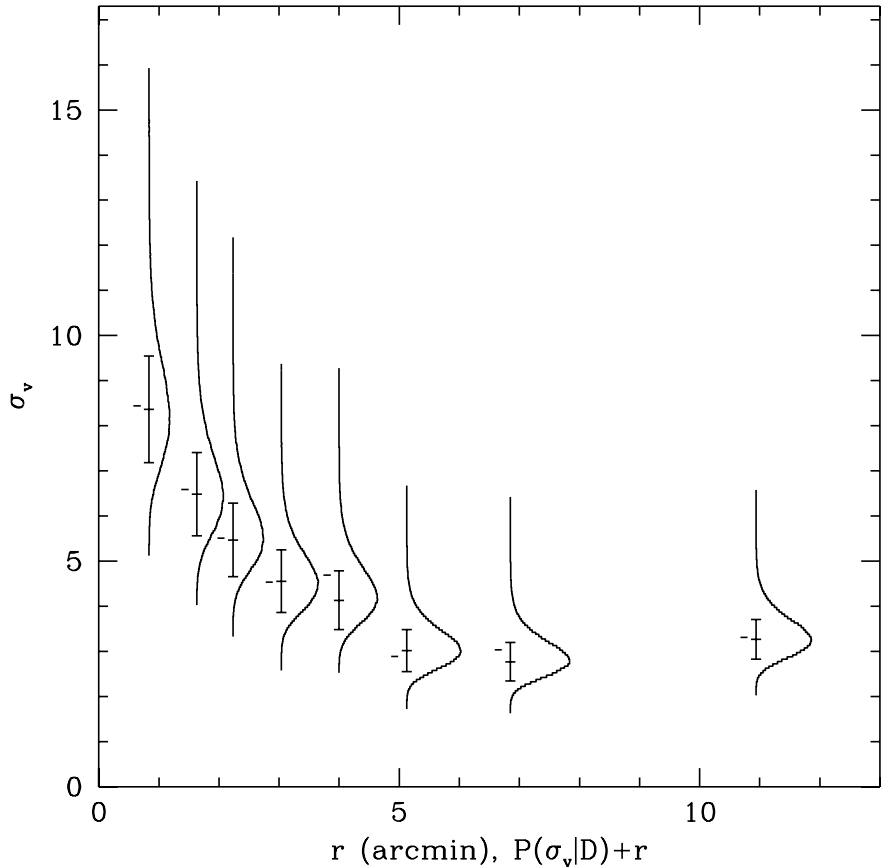}
\caption{As Figure~\protect{\ref{F:Brot}} for the velocity dispersion
profile assuming there is rotation. The dispersion values from
Figure~\protect{\ref{F:dispnorot}} are shown a tick marks offset to
the left of the new points. Note the decrease in the dispersion at
4\arcmin.
\label{F:disprot}}
\end{figure}

If we use the rotating model for the velocities discussed in
\S\ref{S:Bmodels}, assuming a single amplitude and position angle for
the whole sample, then the mean velocity is the same.  We have taken
the hyper-parameter $\delta$ (see \S\ref{S:Bmodels}) to be 5 km
s$^{-1}$.  Rotation is detected.  The amplitude is $1.5\pm 0.4$ km
s$^{-1}$ and the axis of rotation is at $125\pm 19\arcdeg$. The
posterior odds ratio for a rotating model with respect to a
non-rotation model is 15 to 1, indicating that the rotating model is
more likely.  This rotation is displayed in Fig.~\ref{F:pa}. The
corresponding velocity dispersion profile, corrected for rotation, is
shown in Fig.~\ref{F:disprot}. For comparison, the modal values for
the non-rotating analysis are shown as horizontal dashes offset to
slightly smaller radii.  Again, the modes and sizes of the symmetric
error regions are given in Table~\ref{Table:Disp}.  Rotation, if
present, acts to increase the observed velocity dispersion if not
accounted for. The large decrease in the velocity dispersion for the
4\arcmin\ point indicates that it is the stars at this radius that are
most strongly affected by rotation. The dispersion at 7\arcmin\ also
decreases somewhat for the rotating model.  The probability that the
last point is higher than the next-to-last is now 77\%, i.e. the odds
are roughly 3.3 to 1 in favor of the last point being higher, somewhat
higher than for the case without rotation.  For the last point and the
point at 5\arcmin, the probability that the last is larger is now only
65\% giving odds of 1.8 to 1.  The velocity dispersion now appears to
reach its minimum closer to 7\arcmin\ rather than at 5\arcmin\ as in
Fig.~\ref{F:dispnorot}.

Since it has been claimed by Gebhardt et al.\markcite{geb97} (1997)
that the rotation properties change with radius, we have reanalyzed
our sample by looking for rotation in each of our bins separately.  We
have kept the mean velocity fixed at $-106.9$ km s$^{-1}$. The results
are shown in Table~\ref{T:rotation}. Unlike Table~\ref{Table:Disp},
each bin is analyzed separately in this table.  For each bin, as well
as the overall sample analyzed in one bin (the line marked ``All''),
we give the dispersion without rotation and the dispersion, rotation
amplitude, and position angle assuming rotation. The final column
gives the posterior odds ratio in the sense $P(\hbox{\rm
rotation}|D)/P(\hbox{\rm no rotation}|D)$. Values larger than one
indicate the model with rotation is more likely. For the bins where no
rotation is detected, i.e. those with near zero amplitudes, the errors
are the amplitude at which the integrated probability reaches 0.685,
which corresponds to a $1\sigma$ upper limit.  The only bin in which
there is significant rotation detected is the one at 4\arcmin\ as we
expected. The stars in this bin are highlighted in Figure~\ref{F:pa}.
For a different binning, one with six bins alternating with 35 or 36
stars, no single bin shows significant rotation individually, yet the
combined solution has posterior odds in favor of rotation of 33 to 1.

The proceeding results can be explained as a consequence of a subset
 of the observed velocities combining together to give the rotation
 signal. These stars are distributed across many of the bins and,
 depending on the binning, a given radial range may or may not show
 the rotation signal. What the Bayesian posterior odds ratio gives is
 the strength with which one model is preferred over the other. In the
 last mentioned case, the rotating model with six bins is 33 times
 more likely to be a true description of the data than the
 non-rotating model with six bins. Similarly, for the original
 eight-bin model, the rotating model is about 15 times more likely
 than the non-rotating model. These are the same data; why the
 difference? This is a result of the interaction between the
 velocities of the stars in each bin and the dispersion and overall
 rotation in the model. Experiments with selections of radial
 sub-samples show that a group of 32 stars, including all of those in
 the 4\arcmin\ bin above, have a strong probability of rotation. In
 the six-bin model, these stars are divided into two bins and are
 combined with radially adjacent stars which do not support the
 rotation hypothesis as strongly. The individual bins do not show high
 probabilities of rotation, but the entire sample retains the signal.
 A recent Fokker-Planck study of rotating globular clusters shows that
 the rotation velocity should peak at an intermediate radius (Einsel
 \& Spurzem\markcite{es97} 1997). Since the rotation amplitude is
 small, we would expect to detect it only where it is the strongest.

We conclude that our sample supports the view that rotation is present
in M15. This rotation appears strongest near 4\arcmin, but stars from
the entire sample contribute to the signal. The sample is not large
enough to look for radial changes in the rotation amplitude or
position angle in a way independent of the binning.  The ambiguities
introduced by binning, characteristic of such a parametric approach,
suggests that a non-parametric method would be a better way to resolve
this question.  In either case, more data are required.

Even if M15 does rotate, the amplitude is small and little rotation
appears to occur outside 5\arcmin, i.e. in the region where the
velocity dispersion increases. Thus, rotation can not explain the
increase in the velocity dispersion.

\section{Discussion}
Our observations of velocities in M15 indicate that the velocity
dispersion reaches a minimum at a radius of around 7\arcmin\ and then
appears to increase beyond this radius. Even if it does not increase,
and the current data do not unequivocally require an increase, it is
unlikely that the velocity dispersion continues to decrease at the
rate expected for an isolated cluster.  On theoretical grounds, we
would expect the velocity dispersion to decrease with radius as a
power-law for an isolated cluster, $\sigma(r) = \sigma_0 r^{\alpha}$,
with a power-law index close to $\alpha=-0.5$.

To test this we have used the Bayesian algorithm to fit a model
assuming a power-law relation for the velocity dispersion rather than
the step function (i.e. a profile that is constant across each bin)
used above. This model is also discussed in the Appendix.  We have
excluded the 31 stars in the final bin from the fit. For the model
without rotation $\alpha=-0.46\pm 0.08$ and for the model with
rotation $\alpha=-0.48\pm 0.08$.  In both cases $\sigma_0=7.7$. For
the 31 stars in the final bin, we calculated the posterior
probabilities for the single-dispersion and power-law models, and
hence the posterior odds ratio. Both without and with rotation, the
dispersion calculated by the step-function model is favored for the
last bin.  The odds against the power-law model are 165 to 1 without
rotation and 320 to 1 with rotation.

 This finding strongly suggest that there is an external energy source
that heats the outer regions of the cluster and causes the observed
deviation of the velocity dispersion profile from power-law behavior
in the outermost part of the cluster. The most likely, and expected,
energy source is tidal interaction with the galaxy. Whether the
observed heating is due to the general galactic tide, or is due to
shocks involving disk or bulge passages, is impossible to say given
the current state of the observations and models.

But is an energy source necessary?  While rotation has been ruled out,
it could be argued that unorganized, {\em tangential} anisotropy could
give the observed behavior of the velocity dispersion.  This seems
unlikely. Tonry\markcite{t83} (1983) has computed the velocity
dispersion profile for spherical galaxies of varying anisotropy. Even
for extreme tangential anisotropy, there is an increase in the
velocity dispersion only inside of the effective radius of an
$r^{1/4}$ law. Beyond this point, the velocity dispersion decreases,
whatever the degree of anisotropy. Thus it is difficult to see how
tangential anisotropy can increase the velocity dispersion in the
outermost region of a globular cluster.  Further, modeling of isolated
globular clusters clearly demonstrates that the outer parts of such
clusters would be strongly {\em radially} anisotropic
(Larson\markcite{l70} 1970, Spitzer \& Shull\markcite{ss75} 1975,
Cohn\markcite{c79} 1979). To convert this to tangential anisotropy
would require external forces; those originating with the host
galaxy. Thus, the explanation of tangential anisotropy, even if it
could provide an increase in the velocity dispersion, and Tonry's
results suggest otherwise, still requires the influence of the
galactic tidal field to change the orbits of the cluster stars. To
convert radial orbits to circular orbits at the same apocenter
requires energy input, i.e. tidal heating.

The increase in the velocity dispersion that we observe in the outer
part of M15 is qualitatively very similar to that seen in the models
of Allen \& Richstone\markcite{ar88} (1988).  Their results suggest
that the tidal radius should be identified as the location of the
minimum velocity dispersion, at a radius of 7\arcmin, which is very
different from the 23\arcmin\ tidal radius measured by Grillmair et
al.\markcite{grill95} (1995). It is premature, however, to draw firm
conclusions, since the theory of tidal heating and shocking has
advanced since Allen \& Richstone\markcite{ar88} (1988). In
particular, the recognition of the importance of the second order
effects (Kundi\`c \& Ostriker\markcite{ko95} 1995), and the new
appreciation of the limitations of the assumption of adiabatic
invariance (Weinberg\markcite{w94} 1994), require new models. Such
modeling is well within current computational capabilities.  What are
required are models that can be compared with observations. If new
models support the idea that the minimum in the velocity dispersion
marks the edge of the region containing most of the bound stars, then
our results are indeed in contradiction with the analysis of the
surface density profile by Grillmair et al.\markcite{grill96} (1996).

Besides new models, additional observations are also required in order
to improve the statistical significance of our results and to clarify
the role of rotation. We have observed virtually all the stars on the
giant branch brighter than $V=16.6$ outside the central region and
inside 18\arcmin. Below this magnitude, confusion with the galactic
field becomes much stronger. Further out, the fraction of members is
similarly lower. While Hydra is an efficient instrument for observing
large numbers of stars, better selection criteria, based for example
on metallicity discriminating photometry using the Washington or DDO
systems, are required. We have made such observations on the
Washington system for the globular cluster M92, resulting in a
striking increase in the fraction of stars in the sample selected for
Hydra observation proving to be members.  These observations will be
presented in a future paper.

\acknowledgments 
We thank K. Cudworth for the list of stars in his proper-motion
program in advance of publication. We appreciate the support of grants
from the Indiana University College of Arts and Sciences and Office of
Research, and the University Graduate School.

\end{document}